\documentstyle[art12]{article}
\def\({\left(}
\def\){\right(}
\def\[{\left[}
\def\]{\right]}
\begin{document}
\title{
Covariant formulation of non-Abelian gauge theories without
anticommuting variables.} \author{ A.A.Slavnov \thanks{E-mail:$~~$
slavnov@mi.ras.ru} \\ Steklov Mathematical Institute, Russian Academy of
Sciences,\\ Gubkina st.8, GSP-1,117966, Moscow, Russia } \maketitle

\begin{abstract}
A manifestly Lorentz invariant effective action for Yang-Mills theory
depending only on commuting fields is constructed. This action posesses a
bosonic symmetry, which plays a role analogous to the BRST symmetry in the
standard formalism.

\end{abstract}

\section {Introduction}

A peculiar feathure of non-Abelian gauge theories is the appearance of
anticommu- \\ting scalar fields (Faddeev-Popov ghosts) in a Lorentz
covariant local effective action \cite{FP}. It is beleived to be an
unavoidable price to be paid for the construction of a manifestly Lorentz
invariant local effective action.

Locality of the effective action is the essential ingredient of the proof
of unitarity in the Lorentz covariant quantization scheme for nonabelian
gauge fields (BRST quantization) \cite{KO}. However in this approach there
is one problem which, to my knowleadge, has never been
seriously discussed. It is known that Fermi-Dirac quantization
of scalar fields contradicts local commutativity. Therefore it is by no
means evident that a theory including anticommuting scalar fields
respects a causality condition. Of course it is not a serious problem
for the Yang-Mills theory, as in this case one can prove the equivalence
of the formulation based on the Lorentz invariant action and the Coulomb
gauge formulation, in which causality and unitarity is manifest.
However, if one applies BRST quantization to more complicated models,
where a true Hamiltonian formulation is not at hand, this problem deserves
special investigation. For this reason it would be desirable to have a
Lorentz covariant formulation which does not involve anticommuting
scalar fields.

In this paper we shall show that using a bosonization procedure proposed
by us earlier \cite{Sl1}, \cite{Sl2} one can construct an alternative
effective action which includes only commuting fields. This action
describes a five dimensional constrained system, which in the physical
sector is equivalent to the usual Yang-Mills theory. Our bosonic effective
action posesses a symmetry which plays a role similar to the BRST
symmetry \cite{BRS} in the usual approach.In particular this symmetry
generates the relations between the Green functions which are equivalent
to Generalized Ward Identities \cite{Sl3}, \cite{T}.

 \section {Local bosonic effective action.}

Our Lorentz covariant formulation is based on the following effective
action
\begin{eqnarray}
S= \int_{-L}^{L}dt \int d^4x \{ \frac{1}{2L}[- \frac{1}{4}F_{\mu
\nu}^aF_{\mu \nu}^a- \frac{1}{2 \alpha} ( \partial_{\mu}A_{\mu})^2]-
\nonumber \\
-( \frac{i}{2} \frac{ \partial \bar{c}}{ \partial t}+M^+M \bar{c})^ac^a+
h.c.+ \chi^a( \bar{c}^a+c^a) \}
\label{1}
\end{eqnarray}
Here $F_{ \mu \nu}^a$ is the usual Yang-Mills curvature tensor and
\begin{equation}
M= \partial_{\mu}D_{\mu}
\label{2}
\end{equation}
where $D_{\mu}$ is the usual covariant derivative
\begin{equation}
D_{\mu}^{ab}= (\delta^{ab}-gt^{abc}A_{\mu}^c)
\label{3} \end{equation}
The gauge fields $A_{\mu}(x)$ and the Lagrange multiplier $\chi$  depend
 on four dimensional coordinates $x_{\mu}$, whereas the scalar fields
$ \bar{c}(x,t), \quad c(x,t)$ depend also on extra variable $t$, which
acquires values in the interval $ -L \leq t \leq L$. The fields $A_{\mu}$
and $ \chi$ are Hermitean, whereas $ \bar{c}$ and $c$ are conjugate to
each other. All the fields are commuting.

We claim that in the limit $L \rightarrow \infty$ the Yang-Mills field
Green functions generated by the effective action (\ref{1}) coincide
with the corresponding Green functions in the standard Faddeev-Popov
formalism. More precisely
\begin{eqnarray}
Z(J_{\mu})= \lim_{L \rightarrow \infty} \int \exp \{i[S+ \int
d^4xJ_{\mu}^a(x)A_{\mu}^a(x)] \}d \bar{c} dc d \chi dA_{\mu} \nonumber \\
= \exp \{i \int d^4x[- \frac{1}{4}F_{\mu \nu}^aF_{\mu \nu}^a- \frac{1}{2
\alpha}( \partial_{\mu}A_{\mu})^2+J_{\mu}^aA_{\mu}^a] \}
\det{M} dA_{\mu}
\label{4} \end{eqnarray}
where $ \det{M}$ is the usual Faddeev-Popov determinant. The integration
in the eq.(\ref{4}) goes over the fields $ \bar{c}(x,t), \quad c(x,t)$
satisfying the following boundary conditions
 \begin{equation}
c(x,L)=B(x), \quad \bar{c}(x,L)=B^+(x)
\label{5} \end{equation}
where $B(x)$ is an arbitrary, but fixed function, bounded in the limit $L
\rightarrow \infty$. It may depend on other fields, but the simplest
choice is $B=0$. The boundary conditions in fourdimensional coordinates
$x_{\mu}$ are the usual ones (see \cite{SF})

To prove the equation (\ref{4}) we calculate the integral over $ \bar{c},
 c$ explicitely. First of all we note that the operator $M^+M$ is
Hermitean and we can introduce a complete system of it's eigenfunctions
\begin{equation}
M^+M \phi_{\alpha}=m_{\alpha} \phi_{\alpha}
 \label{6} \end{equation}
In this basis the integral over $ \bar{c}, c$ may be written in the form
\begin{equation}
Z_c= \lim_{L \rightarrow \infty} \int \exp \{i \int_{-L}^{L}dt
\sum_{\alpha}-( \frac{i}{2} \frac{ \partial \bar{c}^{\alpha}}{ \partial t}
+ m_{\alpha} \bar{c}^{\alpha})c^{\alpha}+ h.c.+
\chi_{\alpha}(\bar{c}^{\alpha}+c^{\alpha}) \}d \bar{c}^{\alpha}
dc^{\alpha}
\label{7} \end{equation}
Making the change of variables
\begin{eqnarray}
 \bar{c}^{\alpha}(t) \rightarrow \exp \{2im_{\alpha}t \}
\bar{c}^{\alpha}(t)  \nonumber \\
c^{\alpha}(t) \rightarrow \exp \{-2im_{\alpha}t \}c^{ \alpha}(t)
\label{8}
\end{eqnarray}
we can rewrite it in the form
\begin{eqnarray}
Z_c= \lim_{L \rightarrow \infty} \int_{-L}^Ldt \sum_{\alpha}[-
\frac{i}{2}( \frac{ \partial \bar{c}^{\alpha}}{ \partial t}c^{\alpha}-
\bar{c}^{\alpha} \frac{ \partial c^{\alpha}}{ \partial t})+ \nonumber \\
\chi^{\alpha}( \exp \{2im_{\alpha}t \}c^{\alpha}+ \bar{c}^{\alpha} \exp
 \{2im_{\alpha}t \}) d \bar{c} dc d \chi
\label{9}
\end{eqnarray}
The integral over $ \bar{c}, c$ is saturated by the stationary
values of $ \bar{c}, c$, defined by the classical equations:
\begin{eqnarray}
i \frac{ \partial c^{\alpha}}{ \partial t}+ \exp \{2im_{\alpha}t \}
\chi^{\alpha}=0 \nonumber \\
-i \frac{ \partial \bar{c}^{\alpha}}{ \partial t}+ \exp \{-2im_{\alpha}t
\} \chi^{\alpha}=0 \nonumber \\
c^{\alpha}(L)=B^{\alpha} \exp \{-2im_{\alpha}L \}; \quad
\bar{c}^{\alpha}(L)=B^{\alpha} \exp \{2im_{\alpha}L \}
\label{10}
\end{eqnarray}
The solution of these equations is
\begin{eqnarray}
c^{\alpha}= \frac{\chi^{\alpha}}{2m_{\alpha}} \exp \{2im_{\alpha}t
\}+A_{\alpha} \nonumber \\
A_{\alpha}=B_{\alpha} \exp \{-2im_{\alpha}L \}-
\frac{\chi^{\alpha}}{2m_{\alpha}} \exp \{2im_{\alpha}L \} \nonumber \\
 \bar{c}^{\alpha}=(c^{\alpha})^+
\label{11}
\end{eqnarray}
Substituting these solutions into the integral (\ref{9}) and integrating
over $t$ one gets
\begin{equation}
Z_c= \lim_{L \rightarrow \infty} \int \exp \{i \sum_{\alpha}[
\frac{(\chi^{\alpha})^2L}{m_{\alpha}}+ \frac{\chi^{\alpha}}{m_{\alpha}}
\sin(2m_{\alpha}L)(A_{\alpha}+A_{\alpha}^+)] \}d \chi
\label{12} \end{equation}
Renormalizing the fields $\chi^{\alpha}, \quad \chi^{\alpha} \rightarrow
L^{-1/2} \chi^{\alpha}$ and integrating over $ \chi^{\alpha}$ one sees,
that
 \begin{equation}
Z_c= \prod_{\alpha} \sqrt{m_{\alpha}}= \det{M}
\label{13}
\end{equation}
 where we used the fact that $ \det{M}= \det{M^+}$. It completes the proof
 of the eq. (\ref{4}).

\section{Symmetry of the classical action}

In the previous section we proved that in the limit $L \rightarrow \infty$
the Green functions generated by the effective action (\ref{1}) coincide
with the usual Yang-Mills Green func- \\tions in covariant $
\alpha$-gauges. Therefore they satisfy Generalized Ward Identities
\cite{Sl3}, \cite{SF}. So one expects that our effective action posesses
some symmetry which plays a role of BRST symmetry in the usual
formalism. Below we shall show that such a symmetry indeed exists. It
is interesting to note that although our construction reproduces the
usual Yang-Mills theory only in the limit $L \rightarrow \infty$, the
symmetry exists even for a finite $L$. In the limit $L \rightarrow
\infty$ the symmetry transformations simplify considerably.

The symmetry of the action (\ref{1}) can be found by trial and error
method. Firstly one notes that transformation of the gauge field $A_{\mu}$
must leave invariant the Yang-Mills Lagrangian. So it is natural to
start with the special gauge transformation of the fields $A_{\mu},
\bar{c}, c$:
\begin{eqnarray}
 \delta A_{\mu}^a= (D_{\mu} \chi)^a \epsilon \nonumber \\
\delta_1 \bar{c}^a=t^{abd} \bar{c}^b \chi^d \epsilon \nonumber \\
\delta_1c^a=t^{abd}c^b \chi^d \epsilon
\label{14}
\end{eqnarray}
Here $\epsilon$ is a constant infinitesimal parameter. We note that
although transformations (\ref{14}) are  special gauge transformations,
they of course represent a global symmetry.

The transformations (\ref{14}) change the gauge fixing term and quadratic
form for the fields $ \bar{c}, c$
\begin{eqnarray}
\delta_1S= \int d^4x \partial_{\mu} \chi^a(x) \int_{-L}^L dtf_{\mu}^a(x,t)
 \epsilon= \nonumber \\
= \int d^4x \partial_{\mu}( \chi^a F_{\mu}^a(x))- \int d^4x \chi^a
\int_{-L}^L dt \partial_{\mu}f_{\mu}(x,t) \epsilon
\label{15}
\end{eqnarray}
Here the first term at the r.h.s. is a total divergency and the second
term may be compensated by the approptiate variation of fields $ \bar{c},
c$:
\begin{equation}
\delta_2 \bar{c}^a= \delta_2c^a= \frac{1}{2L} \int_{-L}^Ldt
\partial_{\mu}f_{\mu}(x,t) \epsilon
\label{16}
\end{equation}
 The explicit form of the function $f_{\mu}$ is \begin{eqnarray}
f_{\mu}^a= \alpha^{-1}(D_{\mu}( \partial_{ \nu}A_{ \nu})^a+ \nonumber \\
+t^{abd}[(D_{\mu} \bar{c})^b( \partial_{\nu}D_{\nu}c)^a+h.c.] \label{17}
\end{eqnarray}
The transformation (\ref{16}) compensates the variation $ \delta_1S$, but
changes the quadratic form for the fields $\bar{c},c$:
\begin{equation}
\delta_2S=- \frac{i}{2} \int d^4x \delta_2c^a \int_{-L}^Ldt \frac{\partial
\bar{c}^a}{\partial t}+h.c.
+ \int d^4x \frac{i}{2} \delta_2 \bar{c}^aM^+M \int_{-L}^L dt c^a+h.c.
\label{18}
\end{equation}
In this equation the first term is a total derivative over $t$ and the
second term is proportional to
\begin{equation}
\int_{-L}^Ldt( \bar{c}^a(x,t)+c^a(x,t))
\label{19}
\end{equation}
and vanishes at the constraint surface. To provide a manifest invariance
of the action one can eliminate the variation (\ref{18}) by the following
change of Lagrange multiplier $\chi$:
\begin{equation}
\delta \chi^a=- \frac{i}{2} MM^+\delta_2 \bar{c}^a +h.c
\label{20}
\end{equation}
Combining the equations (\ref{14}, \ref{16}, \ref{20}) we get the complete
symmetry transformation of the bosonic effective action (\ref{1}).
These transformations change the effective Lagrangian by a total
derivative and via Neuther theorem lead to the existence of a
five-dimensional conserved current
\begin{equation}
\partial_tj_t+ \partial_{\mu}j_{\mu}=0
\label{21}
\end{equation}
However for physical applications we are interested in the existence of a
conserved charge
\begin{equation}
Q= \int_{-L}^{L}dt \int d^3xj_0
\label{22}
\end{equation}
Integrating the conservation law ( \ref{21}) over $dt$ and $d^3x$ we see
that the integral of $ \partial_ij_i$ vanishes by the usual reasonings due
to fast decreasing of fields at spatial infinity. The integral of $
\partial_tj_t$ gives
\begin{equation}
\int_{-L}^L dt \partial_tj_t=j_t(L)-j_t(-L)
\label{23}
\end{equation}
where
\begin{equation}
j_t= \frac{\delta L}{ \delta \bar{c},_t} \delta \bar{c}- \frac{i}{2}
\bar{c}^a \delta_2c^a+h.c.
\label{24}
\end{equation}
The conservation law (\ref{21}) is fulfilled due to the classical
equations of motion. These equations look as follows \begin{eqnarray} i
 \frac{\partial c}{\partial t}+M^+Mc+ \chi=0 \nonumber \\ -i
\frac{\partial \bar{c}}{\partial t}+M^+M \bar{c}+ \chi=0 \nonumber \\
c(L)=B, \quad \bar{c}(L)=B^+
\label{25}
\end{eqnarray}
It is easy to see that solutions of these equations satisfy the relation
\begin{equation}
\bar{c}(L)=c(-L)
\label{26}
\end{equation}
Due to this property the r.h.s. of eq.(\ref{23}) is zero and
\begin{equation}
\partial_0 \int_{-L}^Ldt \int d^3xj_0(x,t)=0
\label{27}
\end{equation}

For a finite $L$ the symmetry transformations are rather complicated. In
particular they are nonlocal in extra variable $t$. However for physical
applications we are interested in the limiting case $L \rightarrow
\infty$. One can show that in this limit all nonlocal terms give vanishing
contributions to relevant quantities and one can use the asymptotic
symmetry transformations
\begin{eqnarray}
 \delta A_{\mu}^a= (D_{\mu} \chi)^a \epsilon \nonumber \\
\delta \bar{c}^a=[t^{abd} \bar{c}^b \chi^d + \frac{1}{2 \alpha
L}(D_{\nu} \partial_{\nu}(\partial_{\mu}A_{\mu}))]\epsilon \nonumber \\
\delta c^a=[t^{abd}c^b \chi^d + \frac{1}{2 \alpha L}
(D_{\nu} \partial_{\nu}(\partial_{\mu}A_{\mu}))]\epsilon \nonumber \\
\delta \chi^a= \frac{1}{2 \alpha L}[MM^+M( \partial_{\mu}A_{\mu})]^a
\epsilon \label{28} \end{eqnarray} The asymptotic symmetry (\ref{28}) do
not leave the action invariant, but transforms it to another action $S_L
\rightarrow \tilde{S}_L$, which in the limit $L \rightarrow \infty$ leads
to the same physical conclusions.  In the next section we shall show how
using he asymptotic symmetry (\ref{28}) one can derive Generalized Ward
Identities.

\section{Generalized Ward Identities.}

To derive Generalized Ward Identities we consider the following generating
 functional
 \begin{equation}
 Z(J_{\mu}, \eta)= \lim_{L \rightarrow \infty} \int \exp \{iS+
 \int d^4x[J_{\mu}^a(x)A_{\mu}^a(x)+ \eta^a(x) \chi^a(x)] \}dA_{\mu}d
 \bar{c}dc d \chi
 \label{29}
 \end{equation}
 Let us make in this integral the change of variables, which is the
 asymptotic symmetry transformation (\ref{28}). As the change of variables
 does not change the integral, we can put
 \begin{equation}
 \frac{dZ}{d \epsilon}=0
 \label{30}
 \end{equation}
 which is, as we shall see, the equation, generating Generalized Ward
 Identities. Taking into account that the Jacobian of this transformation
 is trivial, we get in this way
 \begin{eqnarray}
 \lim_{L \rightarrow \infty} \int \exp \{iS+
 \int d^4x[J_{ \mu}^a(x)A_{ \mu}^a(x)+\eta^a(x) \chi^a(x)] \}
  \{ \int d^4x [J_{ \mu}^a(x)(D_{
 \mu} \chi(x))^a+ \nonumber \\ + (2aL)^{-1} \eta^a(x)(MM^+M) \partial_{
 \mu}A_{ \mu})^a(x)]+O \} d \bar{c} dc d \chi dA_{ \mu}=0 \label{31}
\end{eqnarray} The term $O=O(1/L)$ arises from noninvariance of the
quadratic form of the fields $ \bar{c}, c$ under asymptotic symmetry
transformations (\ref{28}).  Using the eq.(\ref{16}) one can see that the
variation of the action is quadratic in $ \bar{c}, c$ and proportional to
$L^{-1}$. These terms have to be compared with the quadratic form in the
action (\ref{1}) which is of order $O(1)$. It is easy to verify that these
new terms produce the corrections vanishing in the limit $L \rightarrow
\infty$.

Performing the integration over $ \bar{c}, c$ in the eq.(\ref{31}), we get
\begin{eqnarray}
\lim_{L \rightarrow \infty} \int \exp \{i \int d^4x [- 1/4 F_{ \mu
\nu}^aF_{\mu \nu}^a- 1/(2 \alpha)
(\partial_{ \mu}A_{ \mu})^2+J_{ \mu}^aA_{ \mu}^a+ \eta^a \chi^a-
 \nonumber \\ \chi^a 2L(MM^+)^{-1}_{ab} \chi^b] \}
\int d^4y \{J_{ \mu}^a(y)(D_{ \mu} \chi(y))^a+ \nonumber \\ + (2 \alpha
L)^{-1} \eta^a(y)[(MM^+M) \partial_{ \mu}A_{ \mu}]^a(y) \} dA_{ \mu} d
\chi=0 \label{32} \end{eqnarray}
Differentiating this equation with
respect to $ \eta$ and putting $ \eta=0$ we arrive to the identity
\begin{eqnarray}
\lim_{L \rightarrow \infty} \int \exp \{i[S_{YM}+ \int
d^4x (J_{ \mu}^aA_{ \mu}^a- \chi^a 2L(MM^+)^{-1}_{ab} \chi^b)] \}
\times \nonumber \\
\left[ \int d^4y J_{ \mu}^a(y)(D_{ \mu}
\chi(y))^a \chi^b(z)+ (2 \alpha L)^{-1}\left[(MM^+M)( \partial_{ \mu}A_
{ \mu})\right]^b(z) \right] dA_{ \mu} d \chi=0 \label{33}
\end{eqnarray}
Finally,
performing integration over $ \chi$ and applying the operator
$(MM^+M)^{-1}$ we get the standard Generalized Ward Identity in the form
\cite{Sl3}, \cite{SF}.
\begin{eqnarray}
\int \exp \{i(S_{YM}+ \int d^4xJ_{\mu}^aA_{\mu}^a) \det{M} \times
\nonumber \\
\[ \int d^4y J_{\mu}^b(y)(D_{\mu}^zM^{-1})^{ba}(z,y)+1/ \alpha (
 \partial_{\mu}A_{\mu})^a(y) \]=0
 \label{34}
 \end{eqnarray}
 \section{Discussion}

We showed above that a purely bosonic Lorentz covariant formulation of
Non-Abelian gauge theories is possible provided one introduces the
depenence of auxilliary fields on extra variable. Due to locality of the
effective action and abcense of anticommuting integer spin fields, no
problem with causality arises in this approach. Of course a complete
formulation must also include the discussion of regularization and
renormalization procedure. It can be done but will not be discussed here.
An interesting feathure of our construction is the existence of the
symmetry, which replaces in our approach BRST symmetry. It is important to
note that this symmetry exists even for finite values of $L$ and therefore
may be used for the construction of a Lorentz covariant quantization
procedure alternative to BRST quantization. There are several interesting
problems which remain open in this approach. At present it is not clear
what is a geometrical meaning of the new symmetry. It would be interesting
to try to generalize this construction to supersymmetric theories.

 {\bf Acknowledgements.} \\ This researsh was supported in part by Russian Basic
Research Fund under grant 99-01-00190 and by the president grant for
support of leading scientific schools.$$ ~ $$  
\begin{thebibliography}{99}
{\small \bibitem{FP}L.D.Faddeev, V.N.Popov, Phys.Lett. B25 (1967) 30.
 \bibitem{KO} T.Kugo, I.Ojima, Suppl. Progr.Theor.Phys. 1979, 6.
\bibitem{Sl1} A.A.Slavnov, Phys.Lett. B388 (1996) 147.
 \bibitem{Sl2} A.A.Slavnov, Bosonized formulation of lattice
QCD, hep-th/9611154.
\bibitem{BRS} C.Becci, A.Rouet, R.Stora, Comm.Math.Phys.
42 (1975) 127.
\bibitem{Sl3} A.A.Slavnov, Theor.Math.Phys. 10 (1972) 99.
\bibitem{T} J.G.Taylor, Nucl.Phys. B33 (1971) 436.
 \bibitem{SF} L.D.Faddeev, A.A.Slavnov, Gauge fields.
Introduction to quantum theory. 2nd edition. Addison-Wesley,
1990.} \end {thebibliography}
\end{document}